\documentclass[sigconf,authorversion,nonacm]{acmart}

\AtBeginDocument{%
  \providecommand\BibTeX{{%
    \normalfont B\kern-0.5em{\scshape i\kern-0.25em b}\kern-0.8em\TeX}}}

\usepackage[utf8]{inputenc}
\usepackage{multicol}
\usepackage{subcaption}
\usepackage{graphicx}
\usepackage{amsmath}
\usepackage{natbib}
\usepackage{booktabs}
\usepackage{bbding}
\usepackage{multirow}
\usepackage{tabularx}
\usepackage{comment}
\usepackage{colortbl}

\DeclareMathOperator*{\mode}{mode}

\newcommand\kk[1]{#1}
\newcommand\onobject{\textsc{On-object}}
\newcommand\offobject{\textsc{Off-object}}
\newcommand\combined{\textsc{Combined}}

\title{BeeHIVE: Behavioral Biometric System based on Object Interactions in Smart Environments}

\setcopyright{none}

\begin{document}

\author{Klaudia Krawiecka}
\affiliation{%
  \institution{University of Oxford}
  \city{Oxford}
  \country{United Kingdom}
}
\email{klaudia.krawiecka@cs.ox.ac.uk}

\author{Simon Birnbach}
\affiliation{%
  \institution{University of Oxford}
  \city{Oxford}
  \country{United Kingdom}
}
\email{simon.birnbach@cs.ox.ac.uk}

\author{Simon Eberz}
\affiliation{%
  \institution{University of Oxford}
  \city{Oxford}
  \country{United Kingdom}
}
\email{simon.eberz@cs.ox.ac.uk}

\author{Ivan Martinovic}
\affiliation{%
  \institution{University of Oxford}
  \city{Oxford}
  \country{United Kingdom}
}
\email{ivan.martinovic@cs.ox.ac.uk}

\renewcommand{\shortauthors}{Krawiecka, et al.}

\begin{abstract}
The lack of standard input interfaces in the Internet of Things (IoT) ecosystems presents a challenge in securing such infrastructures.
To tackle this challenge, we introduce a novel behavioral biometric system based on naturally occurring interactions with objects in smart environments.
This biometric leverages existing sensors to authenticate users without requiring any hardware modifications of existing smart home devices.
The system is designed to reduce the need for phone-based authentication mechanisms, on which smart home systems currently rely. It requires the user to approve transactions on their phone only when the user cannot be authenticated with high confidence through their interactions with the smart environment.

We conduct a real-world experiment that involves 13 participants in a company environment, using this experiment to also study mimicry attacks on our proposed system.
We show that this system can provide seamless and unobtrusive authentication while still staying highly resistant to zero-effort, video, and in-person observation-based mimicry attacks.
Even when at most 1\% of the strongest type of mimicry attacks are successful, our system does not require the user to take out their phone to approve legitimate transactions in more than 80\% of cases for a single interaction.
This increases to 92\% of transactions when interactions with more objects are considered.
\end{abstract}

\maketitle

\section{Introduction}

\begin{figure}
  \centering
  \includegraphics[width=\linewidth]{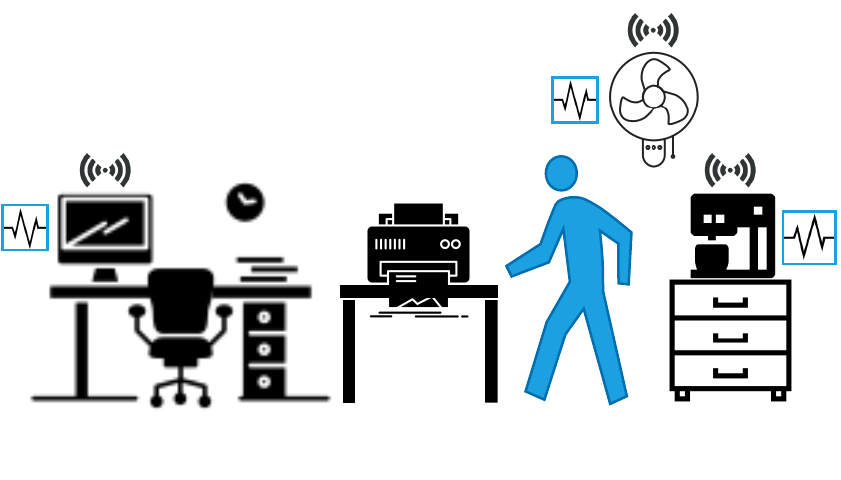}
  \caption{An overview of the \textit{BeeHIVE} system. As the user interacts with the printer, sensors embedded in smart objects surrounding the user and the printer record these interactions. Physical signals generated from the user's movements are picked up by sensors such as accelerometers, pressure sensors and microphones, and are used to profile them. The system authenticates the user before allowing them to perform certain actions, such as payments.}
  \label{fig:sysdesign}
\end{figure}

Projections indicate that by the end of 2021 smart environments will account for over 35\% of all households in North America and over 20\% in Europe~\cite{SOVACOOL2020109663}.
The growing number of smart devices that are incorporated into such environments leads to a wider presence of a variety of sensors.
These sensors can be leveraged to improve the security of smart environments by providing essential input about user activities.
In many environments, the control over specific devices or financial transactions should only be available for an authorized group of users.
For example, smart windows in a child's bedroom should not open when the parent is not present, and the child should not be able to order hundreds of their favorite candy bars using a smart refrigerator. Similarly, not all office workers should have access to a smart printer's history, nor should the visitors in a guesthouse be able to change credentials on smart devices that do not belong to them.
But while there is a need for authentication, smart devices offer limited interfaces for implementing security measures.
This can be mitigated by requiring that the user initiates or approves every transaction through a privileged companion app running on the user's smartphone.
However, this can be very cumbersome as the user needs to have their phone at hand and thus negates many advantages that smart environments offer in the first place.

On-device sensors such as microphones, passive infrared (PIR) sensors, and inertial measurement units (IMUs) have been extensively used to recognize different activities performed by users in the area of Human Activity Recognition (HAR)~\cite{s20010216}. 
Prior work has focused on using one type of input data to authenticate users, such as voice, breath, or gait~\cite{saleema2018voice, meng2018securing, chauhan2018breathing, sun2018accelerometer}. 

In order to make attacks more difficult, several systems have been proposed that rely on diverse types of inputs~\cite{abate2011mubai, 8260821, kim2008multimodal, mavcek2016multimodal, olazabal2019multimodal, gofman2018multimodal}. 
While these approaches are promising, they often do not utilize the full potential of co-located heterogeneous devices in smart environments.
In this paper, we propose the~\textit{BeeHIVE} system that uses sensor data collected during day-to-day interactions with physical objects to implicitly authenticate users without requiring users to change smart home hardware or adapt their behavior.
This system can be used to complement phone-based authentication mechanisms that require users to explicitly approve transactions through privileged apps.
By using \textit{BeeHIVE} in conjunction with a phone-based authentication mechanism as a fallback, smart environments can become more seamless and unobtrusive for users without sacrificing their security.

We conducted a 13-person experiment in a company environment to explore the effectiveness of imitation attacks against our model.
The proposed technique is assessed in three modes of operation to use (1) features from sensors placed on the object with which the user interacts, (2) features only from sensors on co-located objects, and (3) features from both on-device and co-located sensors.
Overall, the outcome of our analysis proves that the system achieves desirable security properties, regardless of the amount of smart office users or the environment configuration.

We make the following contributions in the paper:

\begin{itemize}
    \item We propose a novel biometric based on interactions with physical objects in smart environments.
    \item We collect a separate 13-person dataset in a company setting to study video-based and in-person imitation attacks.
    \item We make all data and code needed to reproduce our results available online.
\end{itemize}

\section{Background and Related Work}
\label{sec:background}

Existing biometric authentication systems that utilize data collected from mobile and smart devices are generally categorized into single-biometric or multi-biometric approaches~\cite{8260821, abate2011mubai}. The systems from the first category collect inputs of a specific type (e.g., sounds, images, acceleration readings) and search for unique patterns. 
On the other hand, multi-biometric systems combine the data extracted from multiple sources to create unique signatures based on different sensor types. They provide more flexibility and lift a number of limitations posed by single-biometric systems, including dependency on certain types of equipment and environmental conditions.
Moreover, they are less prone to mimicry attacks due to the complexity of spoofing multiple modals simultaneously~\cite{yampolskiy2008mimicry}.

\subsection{Single-biometric systems}
The vast majority of existing commercial and non-commercial systems used in smart environment contexts~\cite{barra2013voice, saleema2018voice, meng2018securing, blue20182ma} primarily rely on voice recognition to authenticate users. Since these systems are often vulnerable to voice spoofing and hijacking attacks~\cite{carlini2016hidden, zhang2017dolphinattack, diao2014your, zhang2018understanding}, research efforts shifted towards hardening voice recognition systems by leveraging anti-spoofing mechanisms like proximity detection or second factors~\cite{blue20182ma, meng2018securing}.

With the recent development of new types of Internet of Things and wearable devices, the possibility of using unconventional biometric traits has emerged. For instance, Chauhan et al. observed that microphones can be used to extract breathing acoustics when the user is present in a smart environment~\cite{chauhan2018breathing}.
Similarly, the built-in accelerometers in mobile and IoT devices have been used to characterize gait or human body movements to facilitate authentication~\cite{musale2018lightweight, sun2018accelerometer, musale2019you, batool2017internet}. 
These approaches were the first steps taken to explore the full potential of smart environments to turn contextual and behavioral data into biometric traits for seamless authentication.

\subsection{Multi-biometric systems}

To improve adaptability and reduce the inaccuracy of single-biometric systems, various multi-biometric systems have been proposed~\cite{abate2011mubai, 8260821, kim2008multimodal, mavcek2016multimodal, olazabal2019multimodal, gofman2018multimodal}. 
One approach is to combine two biometric traits to fingerprint users~\cite{mavcek2016multimodal, olazabal2019multimodal, gofman2018multimodal}.
For example, Olazabal et al.~\cite{olazabal2019multimodal} proposed a biometric authentication system for smart environments that uses the feature-level fusion of voice and facial features. 
These solutions, however, still require users to actively participate (e.g. by shaking devices or repeating specific hand wave patterns) in the authentication process and rely on the presence of specific sensors in the smart environment.
To address such limitations, the MUBAI system~\cite{abate2011mubai} employs multiple smart devices to extract various behavioral and contextual features based on well-known biometric traits such as facial features and voice recognition. 

\subsection{Interaction-based biometric systems}
Interaction-based biometric systems have emerged from the observation that physical interactions with devices can uniquely identify users. Such systems have been widely discussed for mobile platforms~\cite{teh2016survey}.
Typically, on-device sensors are employed to measure touch dynamics or user gestures~\cite{teh2016survey, tafreshi2017tiltpass, lee2017secure}.
For example, users can be profiled based on how they pick up their phones or how they hold them~\cite{holdandsign}.
Similar techniques have been used in smart environments; however, most of the existing solutions not only require the user to actively participate in the authentication process but also rely on a specific setup.
Our goal is to introduce a biometric system that continuously and seamlessly authenticates the users while they are interacting with the devices around them without restrictions on sensor placement.

\subsubsection{SenseTribute}
Closest to our work is SenseTribute~\cite{han2018smart}, which performs occupant identification by extracting signals from physical interactions using two on-device sensors---accelerometers and gyroscopes.
Its main objective is to attribute physical activities to specific users.
To cluster such activities, SenseTribute uses supervised and unsupervised learning techniques, and segments and ensembles multiple activities.

There is a palpable risk in real-world smart environments that users will attempt to execute actions that they are not authorized for.
This requires means for not just identification, but also authentication.
Therefore---in contrast to SenseTribute, which focuses on user identification---the main objective of our system is user authentication, for which we conduct a more extensive experiment evaluating various types of active attacks.
In office and home environments, it is easy for anyone to observe interactions made by authorized users, and it is natural that, for example, kids may seek to imitate their parents.
Going beyond previous work, we therefore evaluate the robustness of our system against mimicry attacks based on real-time observation or video recordings.

Furthermore, SenseTribute expects all objects to be equipped with sensors. However, this is not always a realistic assumption, as sensors are often deployed only near (but not on) interaction points. Thus, we propose a system that uses nearby sensors present in co-located IoT devices to authenticate user interactions. 

\section{System Design}
\label{sec:sysmodel}

The heterogeneous nature of smart devices makes it possible to sample different types of user interactions.
The main purpose of this work is to show that such interactions with various objects in smart environments are distinctive and can be used to profile users.
The expansion of smart devices, and hence smart environments, will soon make such methods necessary to quickly authorize certain activities, including payments or management of smart devices.
Figure~\ref{fig:sysdesign} shows an overview of the system design.
The proposed \textit{BeeHIVE}~system is meant to complement existing app-based authentication mechanisms used to secure current smart home platforms.
Our system authenticates the user through their interactions with the smart environment and only requires the user to approve transactions through the app as a fallback if it cannot authenticate the user with confidence itself.
In this way, \textit{BeeHIVE} can be used to reduce the reliance on these app-based authentication mechanisms without compromising on the security of the smart home platform.

\subsection{Design goals}
In order to inform the system design and evaluation methodology, we define the following design goals:

\textbf{Unobtrusiveness.} The system should not require users to perform explicit physical actions for the purpose of authentication nor require them to modify their usual behavior. 

\textbf{Low false accept rate.} As the system is designed to be used alongside app-based authentication, it should prioritize low false accept rates to avoid significantly weakening the security of the overall smart environment system.

\textbf{Low friction.} The system should provide a seamless experience to the user wherever possible. This means that false reject rates should be kept low to reduce the need of falling back on the usual app-based authentication of the underlying smart environment platform. However, this should not come at the cost of higher false accept rates.

\textbf{No restrictions on sensor placement.} The system should use data from existing sensors without making restrictions on their placement or orientation. This ensures that the system can be applied to existing deployments purely through software. In addition, the system should not require sensors on each object but instead use sensors on other nearby devices.

\textbf{Robustness to imitation attacks.} Due to the ease of observation in home environments, the system's error rates should not increase significantly even when subjected to imitation attacks.

\subsection{System model}
\label{sec:goals}
In this work, we consider smart environments where objects such as fridges or cupboards are augmented by smart devices that monitor their state and provide access to enhanced functionality.
People naturally interact with many of these smart objects during their daily activities.
Each activity consists of a set of intermediate tasks.
For instance, to prepare a meal, a user has to walk to the fridge and open it to collect ingredients.
The user then has to walk to the cupboard to pick up the plates.
Behavioral data of these tasks are measured with different types of sensors with which smart devices are frequently equipped.
As some objects might not have any suitable sensors attached to them, we also consider nearby sensors to profile object interactions.
This is particularly true for physical objects without smart capabilities (e.g., cupboards or drawers).
In order to illustrate these different possible deployment settings, we consider three system configurations:
\begin{itemize}
    \item \onobject, where sensors are mounted directly on the object
    \item \offobject, where only co-located sensor data are considered
    \item \combined, which uses sensor data of both the device on the object as well as from co-located devices
\end{itemize}
We use sequences of interactions to increase confidence in system decisions. This way, the user can be better authenticated if they perform several tasks in succession.
As a simplification, we focus on authenticating one user at a time and do not consider multiple users interacting with objects simultaneously.
It is important to note that in our system a failed authentication does not mean that the user is barred from making transactions.
Instead, they can simply not benefit from the seamless authentication provided by our system and are required to use their phone to approve the requested transaction.

\subsection{Deployment scenarios}
The need to authenticate users arises because physical access to smart objects does not imply authorization to use them.
We consider scenarios in which children or visitors may abuse the trust of their parents or hosts to initiate sensitive operations through smart devices that are unwanted by the owners of said devices.
These operations can include making payments to other people, ordering goods online, changing the configuration of smart devices, or accessing sensitive information stored on these devices.
\kk{For example, a child might want to exploit the restocking mechanism of the fridge to order their favorite sweets while a visitor might unintentionally or intentionally access the viewing history of the smart TV and learn intimate details about their hosts.}

Other possible settings include offices where smart devices are accessible to staff and visitors alike. Often these smart devices allow users to complete administrative tasks through them, such as reordering supplies or accessing the print job history of smart printers. But access to this functionality should be restricted to authorized personnel. In these cases, an implicit authentication of the person executing these tasks as done by our system can avoid cumbersome external authentication methods.

\subsection{Adversary model}
An adversary's ($Adv$) main objective is to convince the smart environment that they are a legitimate user ($U_{L}$).
Such a misclassification can result in permitting $Adv$ to execute on-device financial transactions or any other types of sensitive operations on behalf of $U_{L}$.
We assume that $Adv$ has physical access to the environment, but is otherwise an unprivileged user such as a child or a visitor.
Moreover, $Adv$ cannot tamper with the smart devices by, for example, connecting to the debug port to flash the device firmware.
We also assume that smart devices and the user's smartphone are not compromised; thus, they can be considered a reliable data source. Based on these assumptions, we also exclude the possibility of the attacker interrupting the training phase, which could result in the generation of incorrect biometric signatures of authorized users.

In order to achieve their goal, $Adv$ may attempt to mimic the behavior of $U_{L}$ to generate a matching biometric fingerprint.
Successful mimicry attacks on various biometric systems have been previously demonstrated~\cite{khan2018augmented}.
In our scenarios, we consider three types of such attacks: (1) zero-effort attackers who interact with the environment naturally without attempting to change their behavior, (2) in-person attacks in which $Adv$ can observe legitimate users interacting with IoT devices in person, and (3) video-based attacks in which $Adv$ possesses a video recording of the user interacting with the IoT devices in a smart environment.
While in-person attacks give $Adv$ a possibility to inspect $U_{L}$'s interactions more closely and potentially capture more details, recordings can provide additional time to learn $U_{L}$'s behavior.

\section{Experimental Design}
\label{sec:expdesign}

In order to evaluate the feasibility of authenticating users seamlessly based on their interactions with smart devices, we conducted an experiment in a smart office environment with thirteen participants.
This experiment is further used to study attackers that attempt to copy the behavior of the legitimate user to execute mimicry attacks.

\begin{figure}
  \centering
  \includegraphics[width=\linewidth]{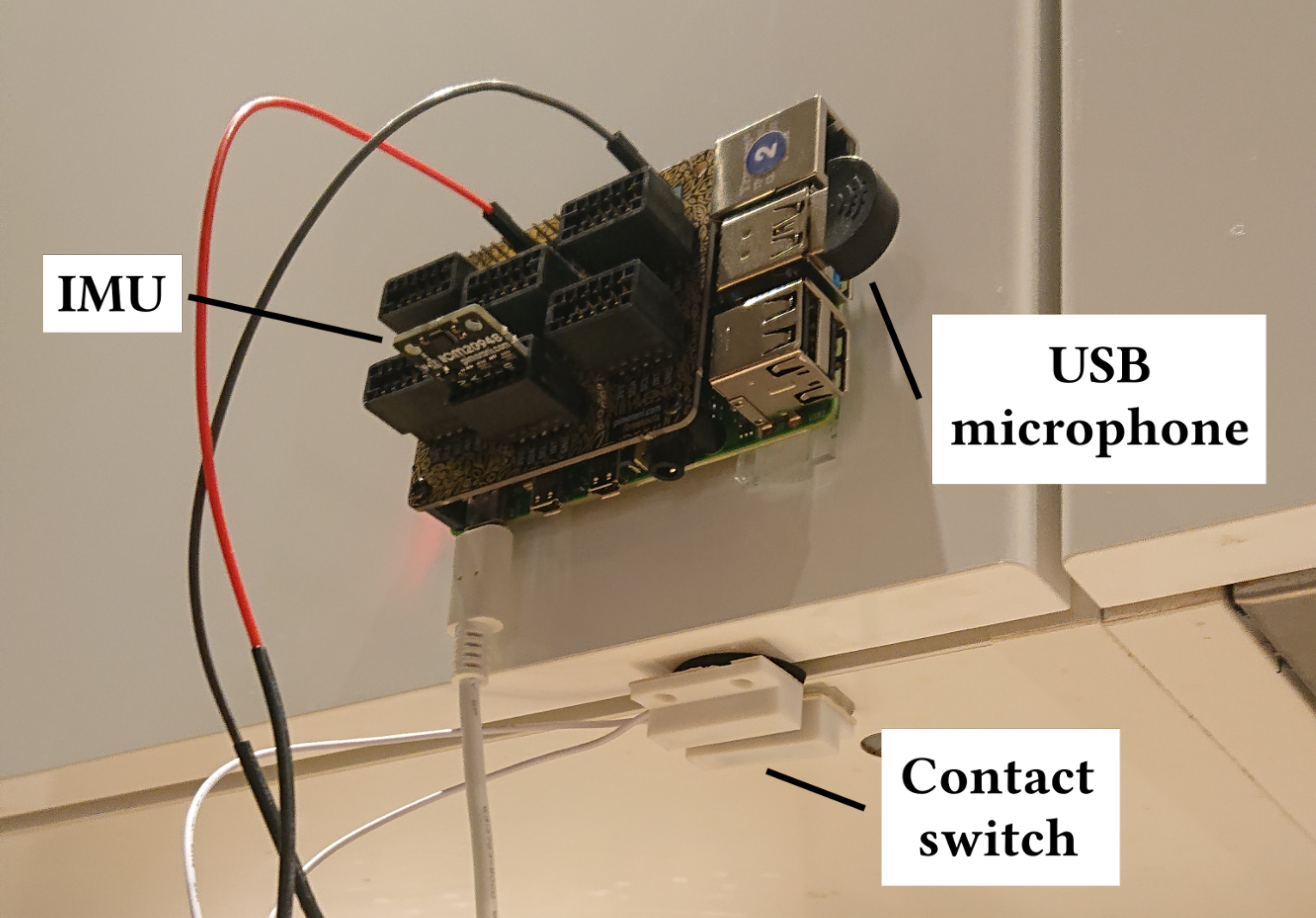}
  \caption{Raspberry Pi attached to a kitchen cupboard. The white magnetic contact switch is used to detect the opening and closing of the cupboard, the USB microphone is used to measure sound pressure levels, and the IMU connected on top of the Raspberry Pi records acceleration, gyroscopic movement and orientation of the interaction.}
  \label{fig:sensorboard}
\end{figure}

\subsection{Data collection}
For our experiment, we collected data from a wide range of typical smart home interactions using sensors similar to those already present in most smart environments.
Since raw sensor data in smart devices are typically inaccessible for developers, we deploy Raspberry Pis equipped with the same types of sensors to simulate such an environment and study object interactions.
We use a total of ten Raspberry Pis equipped with magnetic contact switches, USB microphones (recording sound pressure levels), and ICM20948 IMUs (providing an accelerometer, a gyroscope, and a magnetometer) to collect the data for the experiments.
The Raspberry Pis are fitted to typical home appliances (e.g., fridge or coffee machine) and kitchen furniture (e.g., drawers or cupboards).
The magnetic contact switches are used in place of a typical type of smart office device (i.e., a door/window contact sensor) and they provide the ground truth for the occurrence of interactions with smart objects (e.g., the opening of a kitchen cupboard augmented with a contact sensor).
The IMUs measure the motion sensor data from the interaction (i.e., acceleration, gyroscopic motion, and orientation) and are being polled through the $I^2C$ interface of the Raspberry Pis.
The inputs from the USB microphones are only used to calculate sound pressure levels, but no actual audio data is being stored.
See Figure~\ref{fig:sensorboard} for an example deployment of one of our measurement devices.

The Raspberry Pis are connected to a smartphone running a wireless hotspot.
The data is securely streamed to a remote server and is additionally stored locally on the devices.
A mobile app running on the smartphone is used for labeling and timestamping each run of the experiment and provides time synchronization.

\subsection{Recruitment of participants}

We rented an office space and invited 13 employees of the same company.
We conducted this experiment in adherence to local Covid-19 restrictions and social distancing was observed at all times.
All the participants in our study were compensated for their time and effort.
This project has been reviewed by and received clearance by the responsible research ethics committee at our university, reference number CS\_C1A\_20\_014-1.

\begin{figure}[t]
  \centering
  \includegraphics[width=\linewidth]{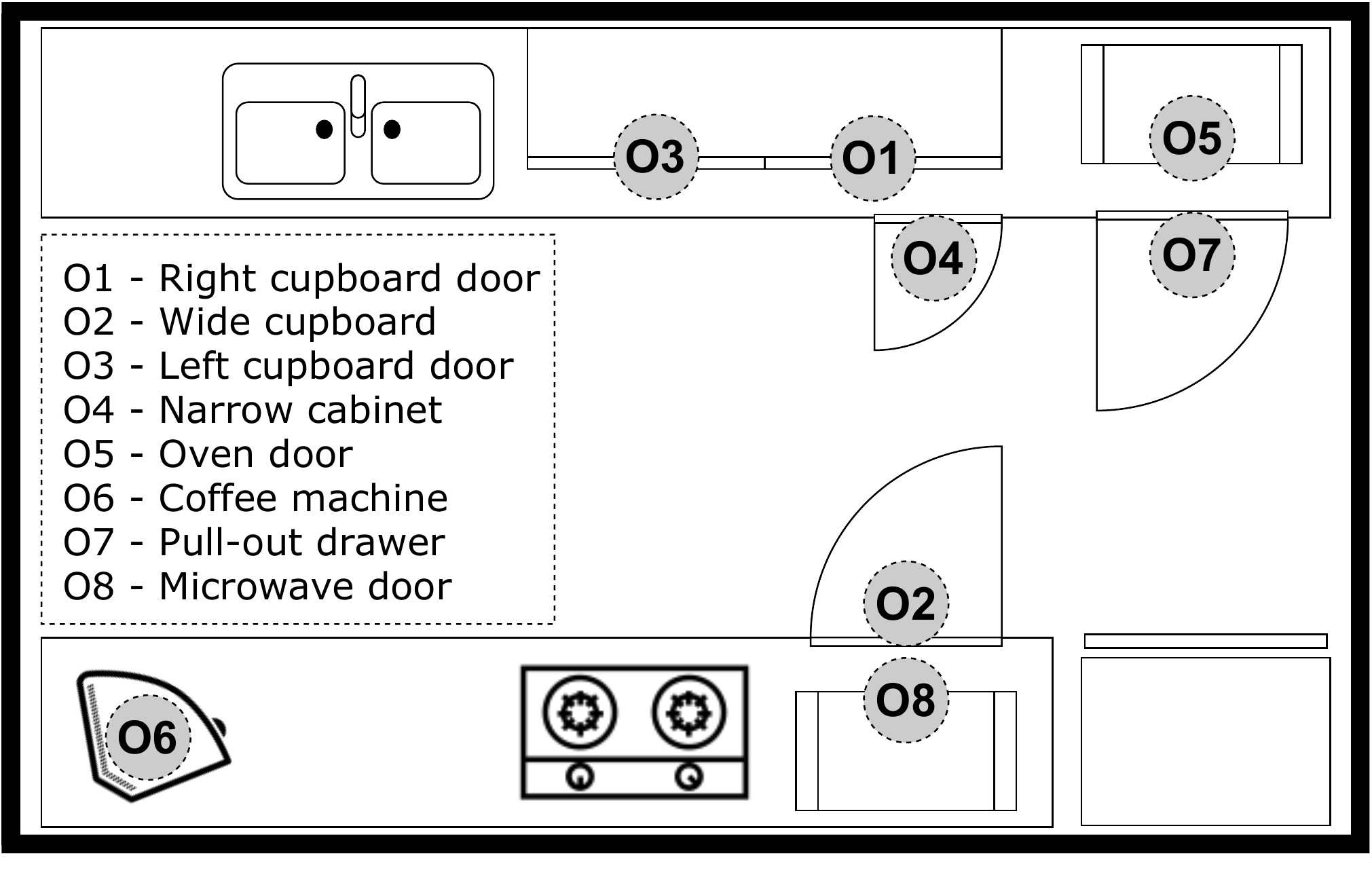}
  \caption{A simplified layout of the room and the arrangement of the objects $O1-O8$ the participants interacted with during the experiment.}
  \label{fig:layout}
\end{figure}

\begin{figure*}[t]
  \centering
  \includegraphics[width=0.76\linewidth]{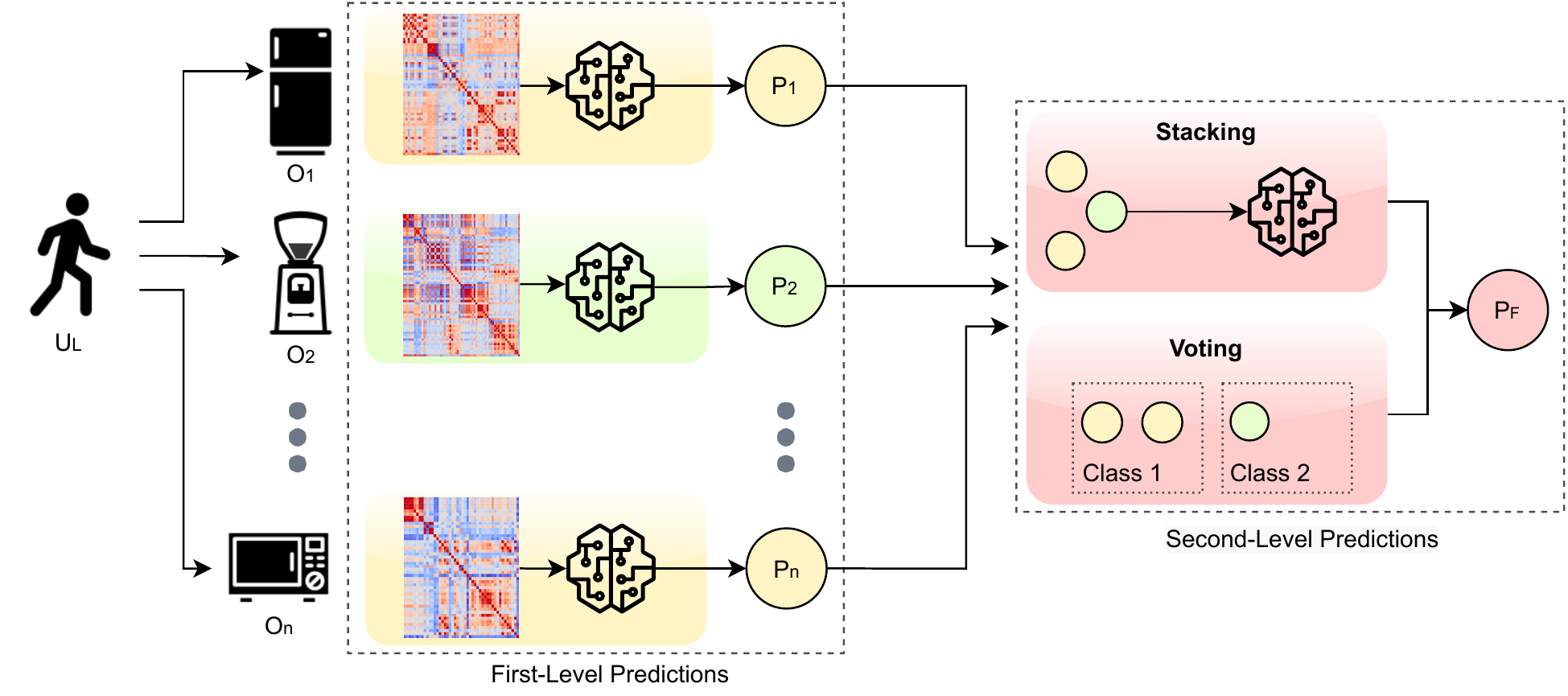}
  \caption{The diagram provides an overview of the processing pipeline of a multi-sensor fusion system. The system extracts relevant features from $U_{L}$'s interactions with objects $O_{1}$ to $O_{n}$ and supplies them to their base-classifiers. Then, the first-level predictions $P_{1}$ to $P_{n}$ are fed into a meta-classifier (i.e., a voting or stacking classifier) that computes the final prediction $P_{F}$.}
  \label{fig:arch}
\end{figure*}

\subsection{Mimicry attacks}

Following Covid-19 regulations, this experiment is conducted remotely in the office kitchen of a hotel company.
In this experiment, we use 8 devices and the device setup is completed by the participants.
They are given a set of our Raspberry Pi sensor boards and they have to set up the devices themselves, according to the provided step-by-step user manual.
An overview of the deployment and the room layout are shown in Figure~\ref{fig:layout}.
As object interactions, we consider in this experiment: 4 cupboards, 1 mini oven, 1 pull-out drawer, 1 microwave, and 1 coffee machine.
Apart from the coffee machine, all of these interactions involve the opening and closing of the doors of the interaction point.
To get the ground truth for the coffee machine interaction, the user first opens a lid on top of the coffee machine which is outfitted with a magnetic contact switch.
The user then proceeds with pressing buttons on the coffee machine, before they end the interaction by closing the lid on top of the machine again.

Each of the participants performs 20 runs of interactions.
Then, one of the participants is randomly chosen as the legitimate user and victim of the attack.
The rest of the participants are split into two groups of six attackers
who can observe the user's interactions with the smart environment.
The first group can only observe the victim in-person, whereas the second group has access to video recordings of previous object interactions which they can study in their own time.

The participants from both groups of attackers then execute the same interactions as the victim, carefully trying to mimic the victim's behavior.
The attackers from the first group (i.e., the in-person group) have to perform this attack on the same day when the observation took place. The participants in the second attack group can watch video recordings of the victim from different angles overnight and only have to execute the attack on the following day.

\section{Methods}
In this paper, we formally define a task $T$ as a physical interaction initiated by the user $U_{L}$ with an object $O$.
Each task can be modeled as a time series $T = \{X_{1}, X_{2}, \dots, X_{z}\}$,
which is constructed from the data collected by on-device sensors, including microphones, accelerometers, gyroscopes, and magnetometers.
The variable $X_{t}$ represents a physical signal generated by the user while they interact with the smart object at time $t$ in form of a vector of sensor values.
Depending on the combination of sensors on the devices, they can collect diverse inputs.
For example, a smart refrigerator equipped with Inertial Measurement Units (IMUs) can collect acceleration values as vectors of $\langle a_{x}, a_{y}, a_{z} \rangle$ when the user opens or closes its door.
However, a smart coffee machine may be equipped with both an IMU and a microphone, which results in collecting more input data.

Figure~\ref{fig:arch} presents the system overview and explains its processing pipeline.
\kk{Base-learners are weak classifiers that are combined to form an ensemble to facilitate the decision-making process.
When the user performs a sequence of tasks $T_{1}$ to $T_{n}$ on several smart objects, the system extracts the features for these tasks from on-object sensors as well as sensors in proximity.}
Next the features become an input to the $n$ base-learners corresponding to those tasks---resulting in predictions $P_{1}$ to $P_{n}$.
These predictions can either indicate a probability that an observed sample belongs to a certain class or a concrete label from the set of labels $L = {L_{1}, L_{2}, \dots, L_{m}}$, \kk{depending on the framework configuration.}
Finally, the meta-learner gathers all predictions made by all the base-learners and decides on the final prediction $P_{F}$ in the second-level prediction layer.
This way, a smart environment can benefit from the heterogeneous character of smart devices and their built-in sensors by performing a decision-level fusion to improve the classification accuracy. 

\begin{figure}
\begin{subfigure}{\linewidth}
  \includegraphics[width=\linewidth]{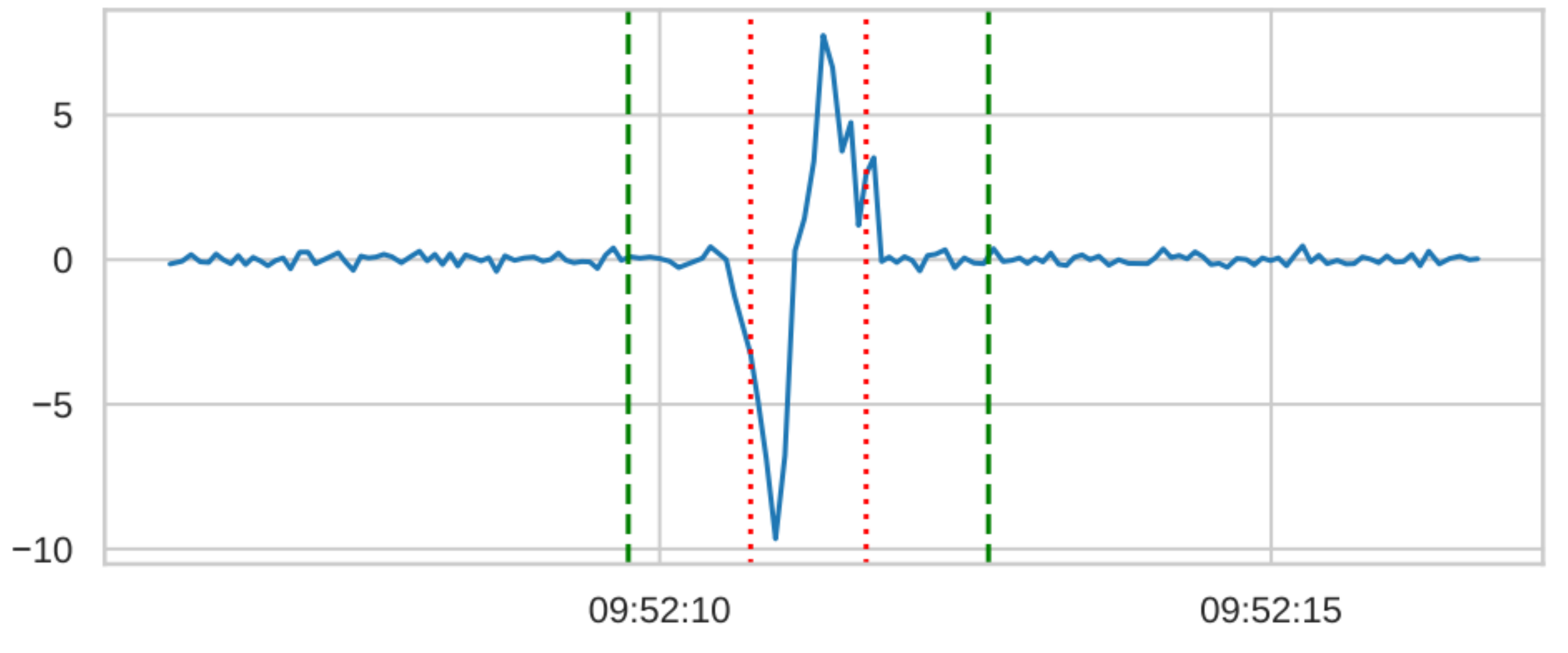}
  \caption{The interaction of the user $U_{1}$ with the narrow cabinet as read by the $x$ axis of a gyroscope.}
  \label{fig:sign_p1} 
\end{subfigure}

\begin{subfigure}{\linewidth}
  \includegraphics[width=\linewidth]{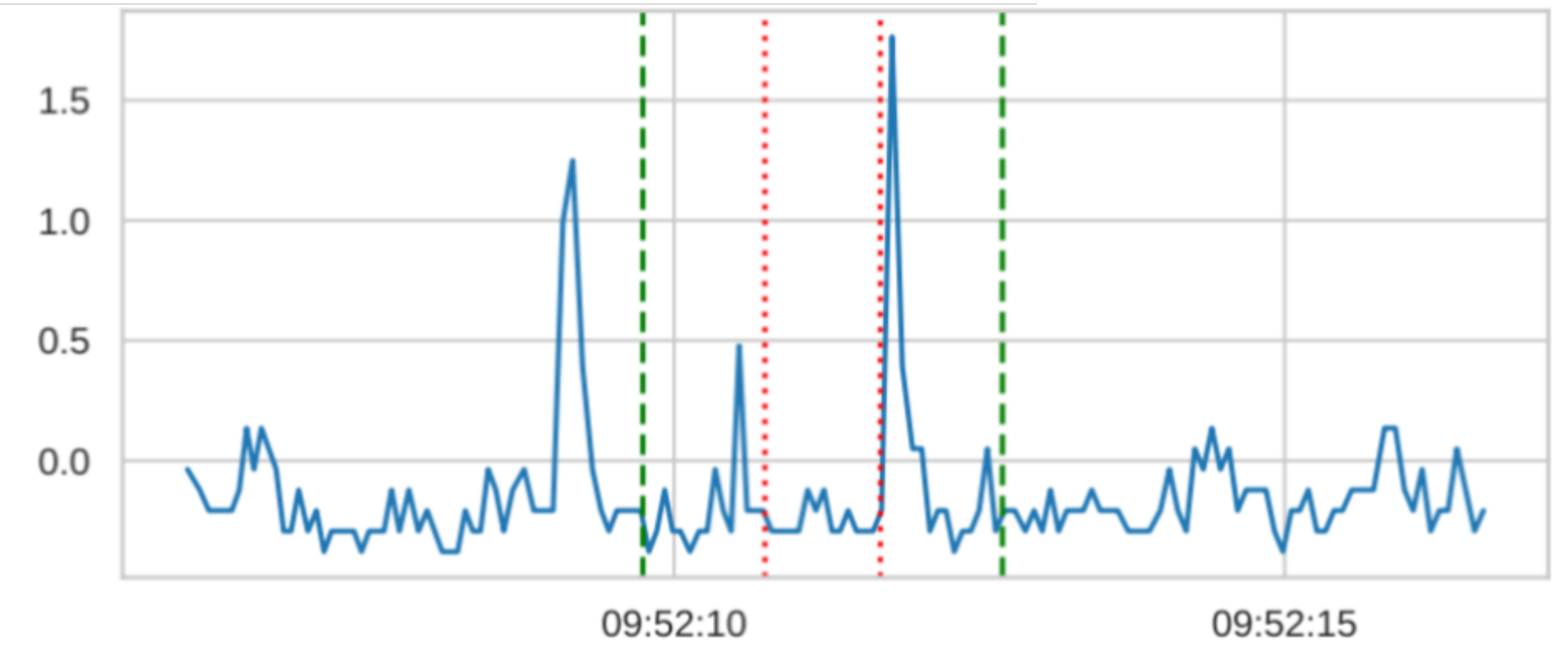}
  \caption{The same interaction of the user $U_{1}$ with the narrow cabinet picked up by the microphone of the co-located wide cupboard $O2$.}
  \label{fig:sign_p2}
\end{subfigure}

\caption{As users interact with smart devices, signals from on-device sensors are collected and processed by the system. Signals (a) and (b) are generated by the participant $U_{1}$ that has interacted with the smart object $O4$.
$O2$ picked up additional input from the same interaction with object $O4$ as they were co-located.
More specifically, these plots illustrate the movements on the $x$ axis of the gyroscope (a) of $O4$ and RMS values (b) picked up by the microphone of $O2$. 
Red dotted lines indicate the start ($t_{0}$) and end ($t_{1}$) of the $T$ task while the green dashed lines denote $\pm$ 1 seconds windows.}
\label{fig:signals}
\end{figure}

\subsection{Preprocessing}

The multitude of $U_{L}$'s interactions in a smart environment translates into physical signals that are received by the sensors of smart devices.
\kk{Figure~\ref{fig:signals} presents the sensor readings when $U_{1}$ interacts with the narrow cabinet during the experiment. While (a) shows the signal that the gyroscope sensor of the cabinet has captured, (b) reveals what has been registered by a co-located sensor.
\kk{Co-located sensors are all sensors in proximity to an object that can capture physical signals originating from interaction with this object.}
The microphone on the wide cupboard recorded two events --- opening and closing the door of the cabinet.}
These movements are part of the task $T$ performed on smart object $O$.
The start and end of $T$ are time-stamped by the contact sensors and denoted as $t_{0}$ and $t_{1}$ respectively (marked with red dotted lines in Figure~\ref{fig:signals}).
The signals from $T$ are segmented by the values of $t_{0} - 1$ and $t_{1} + 1$ before proceeding to the feature extraction phase.
The time-series signals are converted to values characteristic of $h$ sensor types.
As a result, for each $T$, a set of corresponding matrices $S_{1}, S_{2}, ..., S_{h}$ exists that contains vectors of length $m$ of different sensor values $X_{i}$ between $t_{0}$ and $t_{1}$. Thus, with $j \in {1, 2, \dots, h}$ and the sensor component $c_{a}$ where $a \in {1, 2, \dots, m}$, we refer to a single matrix as $S_{j}$ with columns $c_{1}, c_{2} \dots, {c_{m}}$.
The number of columns $m$ for $S_{j}$ is determined by the sensor components (e.g., three axes of a gyroscope sensor $S_{1}$ gives $m_{1} = 3$).
For a smart object with two corresponding sensors, two such matrices will be generated.
The variable $Z_{t_{i}}$ represents a component-specific input value $X_{t_{i}}$ for a column $c_{a}$ generated by a physical signal received by sensor $S_{j}$.
These matrices are then passed as input to the feature extraction function.

\subsection{Feature extraction}

\begin{table}[]
\caption{The decomposition of features that smart devices extract from users' physical interactions.}
\label{tab:features}
\resizebox{\linewidth}{!}{%
\begin{tabular}{ll}
\toprule
Types & Features \\ \midrule
Time-domain features & \begin{tabular}[l]{@{}l@{}}Min, Max, Mean, Median, Std, Var, Kurtosis,\\ Skewness, Shape factor, Absolute energy \\ Mean of central approx. of $2^{nd}$ derivative, \\ Mean/Sum of abs change, Peaks\end{tabular} \\ \midrule
Frequency-domain features & Fourier entropy \\ \bottomrule
\end{tabular}%
}
\end{table}

\kk{For each physical interaction with an object $O_{i}$, the system extracts $h$ matrices with $g  = \sum_{j=1}^{h} m$ columns of time series data segments of particular sensor components of this object and co-located objects.}
Based on these columns, the features are extracted.
More formally, for each smart object $O_{i}$ the system extracts a set of features $F_{O_{i}}$ such that $F_{O_{i}} = \{F_{r} : 0 < r \leq R * g\}$, where:
{\small
\begin{equation}
F_{r} = \{F_{q}([S_{j}]_{a}) : 1 \leq q \leq D\}
\quad for \quad
[S_{j}]_{a} = \begin{pmatrix}
Z_{t_{0}}  \\
\vdots \\
Z_{t_{1}}
\end{pmatrix}
\end{equation}
}
$F_{r}$ represents a set of feature values extracted from a component $a$ (e.g. an $x$-axis of an accelerometer) in $S_{j}$ of length $D$. $R$ denotes the total number of different functions that extract features.

The features retrieved from physical signals are listed in Table~\ref{tab:features}.
\kk{The statistical functions are computed from each column of $S_{j}$, which contains sensor values extracted between $t_{0} - 1$ and $t_{1} + 1$.
We add windows of a second to account for signals that originate from the starting and ending movements.}
These features are categorized into two groups: time-domain features and frequency-domain features.
The majority of the extracted features originate from the time domain because such features are typically well-suited for systems that process large volumes of data due to their low computational complexity.
These features help to analyze the biomechanical effect of a given interaction on physical signals and identify characteristics of movements~\cite{rosati2018comparison}.
For microphone data, we extract sound pressure levels (SPLs) instead of actual audio recordings. 
Thus, statistical functions are applied to SPL values.

\subsection{Feature selection}
\kk{For each smart object, the system selects a subset of extracted features using a filtering method.
This method focuses on verifying whether features are relevant by analyzing their association with the target variable.
The univariate feature selection method used in this work relies on statistical tests to investigate the relationship between variables.}
The features selected from various
sensors are aggregated and become an input to an object-specific base-classifier.
\kk{This is a necessary step in our framework to improve the model performance as well as reduce the computational complexity.}

\subsubsection{Mutual information}
\kk{Mutual information (MI) is used to examine the distinctiveness of a set of features and to test the null hypothesis ($H_{0}$) that negates the existence of a relationship between a feature and an associated target variable.
This method can capture statistical dependencies between variables, explaining whether one variable can provide relevant information about the other one~\cite{beraha2019feature}.
By accepting $H_{0}$, we assume that the extracted feature is not relevant, indicating that it is independent of the target variable.
On the other hand, rejecting $H_{0}$ suggests that the variables can be dependent, so the feature should be considered relevant.
In practice, the null hypothesis is rejected or accepted after examining the resulting non-negative MI scores.
The higher the score, the more significant the feature may be.
The zero score indicates independence between the variables.
As there can be many relevant features, the system has been configured to choose only the top 20 of such features ranked by the highest scores for each object.}

\subsection{Multi-sensor fusion}
As stated in Section~\ref{sec:goals}, we explore and evaluate the potential of heterogeneous smart environments to authenticate users while they carry out their daily activities.
Every node in a smart environment extracts different sets of characteristics from user interactions due to their placement, purpose, and composition of built-in sensors.
Various fusion approaches exist that can boost the detection accuracy and system effectiveness in multi-sensor environments~\cite{aguileta2019multi}. 
Among these fusion techniques, we focus on \textit{decision-level methods} which allow the introduction of multiple classifiers, \textit{base-learners}, that independently undertake a classification task.
This gives a certain degree of autonomy to individual base-learners trained on specific smart object interactions.
Moreover, such an approach allows us to select the most effective feature sets, classifiers and their hyper-parameters for each of the base-learners.
As shown in Figure~\ref{fig:arch}, after each first-level base-classifier makes a prediction, the second-level meta-classifier determines the final outcome.
The efficiency and effectiveness of various fusion techniques at the decision level have been extensively studied in the area of Human Activity Recognition (HAR)~\cite{aguileta2019multi}.
While our focus is on user authentication, we hypothesize that similar approaches can be just as effective in our case.
As such, we compare two ensemble learning techniques that use fundamentally different classification methods but show promise for good performance in our multi-user smart environment scenarios.

\subsubsection{Stacking}

A meta-learner is trained using labels obtained from the first-layer base-learners, as its features~\cite{wolpert1992stacked}.
We chose stacking as a method of linking heterogeneous classifiers since it typically achieves high accuracy and introduces less variance than other approaches.
Combining a multitude of smart objects and their classifiers can be helpful because some of these interactions can classify certain users better than others.
The optimal parameters of the meta-classifier are determined during the training phase by using cross-validation on the training dataset to avoid overfitting.
Based on the combinations of predictions that the meta-learner receives from the base-classifiers, it computes the most accurate label.
Stacking allows combining various classifiers (e.g., k-Nearest Neighbours, Random Forests, Decision Trees, etc.) using different sets of features for each.
In our scenario, the biggest advantage of this approach is that the meta-classifier learns which object interactions predict labels more accurately.
For instance, after a training phase, a meta-classifier learned that tasks performed on $O_{1}$ or $O_{3}$ are more effective in recognizing $U_{L}$ than $O_{2}$.
Therefore, it will account for it in the future while making predictions.
The varying classification effectiveness of individual object interactions and their sensors was a major factor in deciding whether to include the ensemble learning methods in our system. 

\subsubsection{Voting}
\label{sec:voting}
Voting is another ensemble learning method discussed in this paper.
In comparison to stacking, this technique does not require a separate machine learning model to make final predictions.
Instead, it uses the deterministic hard voting algorithm to compute the result.
In our scenario, voting simply indicates that the most reported class label from the set of predictions will be selected as the outcome $P_{F}$.
\begin{equation}
P_{F} = \mode(P)  
\end{equation}
For example, let's assume that there is a set of predictions $P = \{P_{1}, P_{2}, P_{3}\}$ computed for three smart object interactions. For our first object, a smart fridge, the system computed $P_{1}$, which resulted in label $L_{1}$ belonging to $U_{L}$. Then, a smart microwave did not recognize $U_{L}$. But for the coffee machine the system calculated $P_{3}$ that again pointed to $L_{1}$. Thus, the final prediction would result in $L_{1}$.
We decided to assign uniform weights to all base-classifiers and test how different combinations of such classifiers affect the performance of the ensemble system.

\begin{table}[]
    \caption{Search space for Random Forest (RF) hyperparameters. As each base classifier choses their own parameters, the optimal values given here are the most commonly chosen ones.}
    \label{tab:hyper_rf}
    \resizebox{\linewidth}{!}{
\begin{tabular}{lll}
\toprule
Parameter            & Search space                     & Optimal value  \\
\midrule
Number of estimators & 10, 50, 100, 200                 & 100            \\
Tree depth           & 2, 4, 5, 6, 7, 8                 & 7              \\
Number of features   & $\sqrt{N_{F}}$, $\log N_{F}$ & $\sqrt{N_{F}}$    \\ 
\bottomrule
\end{tabular}
}
\end{table}

\begin{table}[]
    \caption{Search space for Support-Vector Machine (SVM) hyperparameters. As each base classifier choses their own parameters, the optimal values given here are the most commonly chosen ones.}
    \label{tab:hyper_svm}
\resizebox{\linewidth}{!}{
\begin{tabular}{lll}
\toprule
Parameter            & Search space                     & Optimal value  \\
\midrule
C & 0.1, 1, 10, 100                 & 0.1            \\
$\gamma$           & 1., 0.1, 0.01, 0.001                 & 0.01              \\
Kernel function   & linear, polynomial, rbf, sigmoid & rbf    \\ 
\bottomrule
\end{tabular}
}
\end{table}

\subsection{Classification tasks}
In this paper, we discuss a supervised learning task that focuses on binary classification to evaluate authentication performance of our system.
We extract features for each base classifier in three different system configurations, namely \onobject{}, \offobject{}, and \combined{}.
These base classifiers create the first-level predictions $P_{1}$ to $P_{n}$ and on their basis, the meta-classifier generates the second-level prediction $P_{F}$.
Since both the Support Vector Machine (SVM) and Random Forest (RF) models tend to outperform other classifiers in HAR tasks~\cite{GARCIACEJA201845}, we chose them as the base classifiers in our task.
For each object interaction classifier, a grid search is performed to find the optimal set of hyper-parameters.
More details can be found in Tables~\ref{tab:hyper_rf} and~\ref{tab:hyper_svm}.
For all SVM-based base-learners, the selected features are first standardized.
Models are trained and tested using 10-fold cross-validation to avoid information leakage.
The resulting classification accuracy is averaged over different folds and used to select the best models.
For stacking, as the second layer model, we chose Logistic Regression due to its simplicity and ease of interpretation~\cite{van_Loon_2020}.

\section{Evaluation}
\label{sec:eval}

\subsection{Distinctiveness of sensor features}

\begin{table}[]
\caption{
Aggregated maximum values of RMI in percentages for the \onobject{} configuration, given different types of on-device sensors.}
\label{tab:rmi_mainonly}
\resizebox{\linewidth}{!}{
\begin{tabular}{lcccc}
\toprule
Object Type &                                     ACC &                                     MAG &                                    GYRO &                                     MIC \\
\midrule
Right cupboard door & \cellcolor[rgb]{0.92, 0.92, 0.92} 31.60 & \cellcolor[rgb]{0.89, 0.89, 0.89} 40.58 & \cellcolor[rgb]{0.75, 0.75, 0.75} 72.32 & \cellcolor[rgb]{0.95, 0.95, 0.95} 20.72 \\
Wide cupboard & \cellcolor[rgb]{0.85, 0.85, 0.85} 50.81 & \cellcolor[rgb]{0.86, 0.86, 0.86} 47.30 & \cellcolor[rgb]{0.79, 0.79, 0.79} 64.45 & \cellcolor[rgb]{0.94, 0.94, 0.94} 25.37 \\
Left cupboard door & \cellcolor[rgb]{0.80, 0.80, 0.80} 62.90 & \cellcolor[rgb]{0.89, 0.89, 0.89} 39.50 & \cellcolor[rgb]{0.75, 0.75, 0.75} 73.90 &  \cellcolor[rgb]{0.98, 0.98, 0.98} 8.09 \\
Narrow cabinet & \cellcolor[rgb]{0.92, 0.92, 0.92} 30.39 & \cellcolor[rgb]{0.91, 0.91, 0.91} 33.28 & \cellcolor[rgb]{0.96, 0.96, 0.96} 19.41 & \cellcolor[rgb]{0.96, 0.96, 0.96} 15.09 \\
Oven door & \cellcolor[rgb]{0.92, 0.92, 0.92} 32.24 & \cellcolor[rgb]{0.79, 0.79, 0.79} 64.46 & \cellcolor[rgb]{0.86, 0.86, 0.86} 49.20 & \cellcolor[rgb]{0.96, 0.96, 0.96} 16.30 \\
Coffee machine & \cellcolor[rgb]{0.91, 0.91, 0.91} 34.07 & \cellcolor[rgb]{0.88, 0.88, 0.88} 41.64 & \cellcolor[rgb]{0.81, 0.81, 0.81} 60.48 & \cellcolor[rgb]{0.88, 0.88, 0.88} 41.67 \\
Pull-out drawer & \cellcolor[rgb]{0.90, 0.90, 0.90} 35.19 & \cellcolor[rgb]{0.94, 0.94, 0.94} 27.21 & \cellcolor[rgb]{0.96, 0.96, 0.96} 19.28 & \cellcolor[rgb]{0.87, 0.87, 0.87} 44.63 \\
Microwave door & \cellcolor[rgb]{0.90, 0.90, 0.90} 35.93 & \cellcolor[rgb]{0.83, 0.83, 0.83} 54.04 & \cellcolor[rgb]{0.88, 0.88, 0.88} 41.12 & \cellcolor[rgb]{0.96, 0.96, 0.96} 15.10 \\
\bottomrule
\end{tabular}
}
\end{table}

\begin{table}[]
\caption{
Aggregated maximum values of RMI in percentages for the \offobject{} configuration, given different types of co-located sensors.}
\label{tab:rmi_collocated}
\resizebox{\linewidth}{!}{
\begin{tabular}{lcccc}
\toprule
Object Type &                                     ACC &                                     MAG &                                    GYRO &                                     MIC \\
\midrule
Right cupboard door & \cellcolor[rgb]{0.72, 0.72, 0.72} 78.63 & \cellcolor[rgb]{0.76, 0.76, 0.76} 71.49 & \cellcolor[rgb]{0.75, 0.75, 0.75} 72.08 & \cellcolor[rgb]{0.96, 0.96, 0.96} 17.33 \\
Wide cupboard & \cellcolor[rgb]{0.68, 0.68, 0.68} 85.86 & \cellcolor[rgb]{0.84, 0.84, 0.84} 53.72 & \cellcolor[rgb]{0.81, 0.81, 0.81} 59.10 & \cellcolor[rgb]{0.93, 0.93, 0.93} 28.07 \\
Left cupboard door & \cellcolor[rgb]{0.72, 0.72, 0.72} 77.69 & \cellcolor[rgb]{0.74, 0.74, 0.74} 75.53 & \cellcolor[rgb]{0.75, 0.75, 0.75} 74.21 & \cellcolor[rgb]{0.96, 0.96, 0.96} 17.49 \\
Narrow cabinet & \cellcolor[rgb]{0.73, 0.73, 0.73} 76.23 & \cellcolor[rgb]{0.75, 0.75, 0.75} 73.43 & \cellcolor[rgb]{0.80, 0.80, 0.80} 61.28 & \cellcolor[rgb]{0.95, 0.95, 0.95} 23.31 \\
Oven door & \cellcolor[rgb]{0.74, 0.74, 0.74} 74.67 & \cellcolor[rgb]{0.75, 0.75, 0.75} 73.52 & \cellcolor[rgb]{0.81, 0.81, 0.81} 60.76 & \cellcolor[rgb]{0.96, 0.96, 0.96} 17.93 \\
Coffee machine & \cellcolor[rgb]{0.60, 0.60, 0.60} 98.19 & \cellcolor[rgb]{0.75, 0.75, 0.75} 74.01 & \cellcolor[rgb]{0.65, 0.65, 0.65} 90.41 & \cellcolor[rgb]{0.88, 0.88, 0.88} 42.96 \\
Pull-out drawer & \cellcolor[rgb]{0.71, 0.71, 0.71} 80.61 & \cellcolor[rgb]{0.77, 0.77, 0.77} 69.31 & \cellcolor[rgb]{0.82, 0.82, 0.82} 56.81 & \cellcolor[rgb]{0.94, 0.94, 0.94} 25.50 \\
Microwave door & \cellcolor[rgb]{0.67, 0.67, 0.67} 86.09 & \cellcolor[rgb]{0.83, 0.83, 0.83} 54.78 & \cellcolor[rgb]{0.82, 0.82, 0.82} 58.37 & \cellcolor[rgb]{0.95, 0.95, 0.95} 20.33 \\
\bottomrule
\end{tabular}
}
\end{table}

\kk{In order to judge the distinctiveness of features by different types of sensors, we use relative mutual information (RMI).
RMI is defined~as
\[\mathrm{RMI}(user,F)=\frac{H(user)-H(user|F)}{H(user)}\]
where H(A) is the entropy of A and H(A$|$B) denotes the entropy of A conditioned on B. Here, $user$ denotes the ground truth of the user performing the object interaction, whereas $F$ is the vector of extracted features.}

Tables~\ref{tab:rmi_mainonly} and ~\ref{tab:rmi_collocated} show the RMI for individual sensors that have been placed on household objects as part of our experiment. These scores represent aggregated maximum values of RMI for a particular sensor on a specific household object $O_{i}$, given different configurations of the system.
Each of these objects introduces a different way for a user to interact with the smart environment. 
Analysis of the distinctiveness of the features extracted from these sensors allows us to understand which ones contribute to better classification performance for a specific type of interaction.
Each device has been equipped with an accelerometer (ACC), a magnetometer (MAG), a gyroscope (GYRO), and a microphone (MIC). 
Generally, we observe that the features extracted from GYRO and ACC exhibit high distinctiveness for most of the interaction types.
\kk{For \onobject, the most distinctive features originate from GYRO whereas for \offobject, ACC appears to supply the most distinctive features.
We observe that, in many cases, the inputs from co-located objects generate higher RMI scores.} 
On the other hand, the features extracted from MIC appear to have relatively low distinctiveness in comparison to other attributes for the majority of interactions.

\kk{Despite its generally low distinctiveness for most interactions, MIC achieves higher RMI values for interactions with the pull-out drawer and is the second most distinctive sensor for the coffee machine when we consider features extracted only from its on-device sensors.}
This can be explained as the drawer's contents make sounds continuously, changing based on how far extended the drawer is,
whereas for most other events the main sounds were caused by the closing of doors---with little difference between users.
Pressing the buttons of the coffee machine on the other hand makes faint sounds which differ between users with regards to the timing of the button presses.

GYRO shows particularly high distinctiveness for most interactions for \onobject, with the exceptions of the narrow cabinet and the pull-out drawer.
The cabinet used in the experiment has a very stiff door that leads to abrupt openings with little variation between users.
\kk{While this reduces the effectiveness of the recognition of users by sensors directly placed on the cabinet, such abrupt openings allow co-located sensors to capture stronger vibrations, hence, provide more accurate distinction.}
The lower RMI values for GYRO for the pull-out drawer can be explained by a lack of rotational movement. 
\kk{Instead, the most distinctive movement characteristics are the sounds and the acceleration which is why MIC and ACC are the most distinctive sensor types for this interaction.}

\begin{table*}[t]
\centering
\caption{\kk{False Reject Rates (FRRs) for interactions with different types of objects in respect to three kinds of attacks given different FAR thresholds. The \onobject{} column presents FRRs for the model with features extracted only from on-device sensors. \offobject{} shows FRRs considering only features from co-located sensors, whereas \combined{} reveals FRRs for the model that uses the combined features from the co-located and on-device sensors. The results are averaged across all smart objects in our experiment.}}
\label{tab:frratfar}
\resizebox{\textwidth}{!}{%
\begin{tabular}{lccc c ccc c ccc}
\toprule
\multirow{2}{*}{\textbf{FAR}} & \multicolumn{4}{c}{\textbf{\onobject{} FRR}} & \multicolumn{4}{c}{\textbf{\offobject{} FRR}} & \multicolumn{3}{c}{\textbf{\combined{} FRR}} \\ 
\cline{2-4} \cline{6-8} \cline{10-12} 
 & Zero-effort &  \multicolumn{1}{c}{Video} & \multicolumn{1}{c}{In-person} && 
 Zero-effort & \multicolumn{1}{c}{Video} & \multicolumn{1}{c}{In-person} && 
 Zero-effort & \multicolumn{1}{c}{Video} & \multicolumn{1}{c}{In-person} \\ \midrule
10\% & 0.1375 & 0.3708 & 0.2250 && 0.0063 & 0.1750 & 0.1250 && 0.0125 & 0.1938 & 0.1500 \\
1\% & 0.2486 & 0.5812 & 0.5087 && 0.0250 & 0.1938 & 0.1500 && 0.0250 & 0.2500 & 0.2875 \\
\bottomrule
\end{tabular}
}
\end{table*}

\begin{table*}[t]
\centering
\caption{\kk{FRRs at two distinctive FAR thresholds for interactions with different types of objects in respect to zero-effort attacks given \onobject{} and \offobject{} configurations. These configurations are compared to emphasize the improvement offered by considering co-located sensors. Presented results are averaged across all users being considered a victim.}}
\label{tab:frratfar_off_on}
\resizebox{0.9\textwidth}{!}{%
\begin{tabular}{lcc c cc c cc}
\toprule
\multirow{2}{*}{\textbf{Object Type}} & \multicolumn{3}{c}{\pmb{$FAR = 10\%$}} & \multicolumn{2}{c}{\pmb{$FAR = 1\%$}}  \\ 
\cline{2-3} \cline{5-6} 
 & \onobject{} FRR &  \multicolumn{1}{c}{ \offobject{} FRR} && 
   \onobject{} FRR & \multicolumn{1}{c}{ \offobject{} FRR} 
 \\ \midrule
Right cupboard door & 0.0526  & 0.0039 && 0.1401 & 0.0154  \\
Wide cupboard & 0.0577  & 0.000 && 0.2231 & 0.0077  \\
Left cupboard door & 0.0369 & 0.0039  && 0.1077 &  0.0039  \\
Narrow cabinet & 0.1141 & 0.0 && 0.2577 &  0.0039  \\
Oven door &  0.0305 & 0.0 && 0.1020 &  0.0  \\
Coffee machine & 0.0154 &  0.0 && 0.0731 & 0.0  \\
Pull-out drawer &  0.0385 &  0.0 && 0.1180 &  0.0  \\
Microwave door & 0.0987 & 0.0115 && 0.2962 &  0.0192 \\
\bottomrule
\end{tabular}
}
\end{table*}

\kk{ACC appears to provide the most distinctive features captured by co-located sensors.
Interestingly, the vibration signals picked up by the co-located sensors exhibit the highest feature distinctiveness during interactions with the coffee maker.
Overall, we notice that \offobject{} features provide better distinctiveness than the features gathered only by \onobject{} sensors.
This suggests that the system can accurately authenticate users by their interactions with objects that do not have sensors directly placed on them.}

\subsection{Authentication performance}

In our experiment, we focus on analyzing the system performance against three types of attacks.
The first part of the dataset contains the samples from the victim as well as zero-effort attack samples from each of the remaining 12 participants.
This dataset is split using 10-fold cross-validation. 
Each test fold is used to evaluate a group of zero-effort attacks since it contains the samples of attackers' regular interactions with objects.
The remaining attack samples are supplied to the zero-effort attack-trained classifier.
To compare and evaluate the effectiveness of different types of attacks on the environment, we report False Reject Rates (FRRs) at different thresholds of False Acceptance Rates (FARs).
The FAR metric allows us to determine how many attempts the attacker was successful in.
On the other hand, FRR specifies how many legitimate samples from a victim have been misclassified as an attack.
Note, that rather than completely preventing the user from executing a transaction this merely means that the user will have to approve the transaction explicitly through their phone.

\begin{figure*}[t]
\centering
\begin{subfigure}[b]{0.33\textwidth}
    \includegraphics[width=\linewidth]{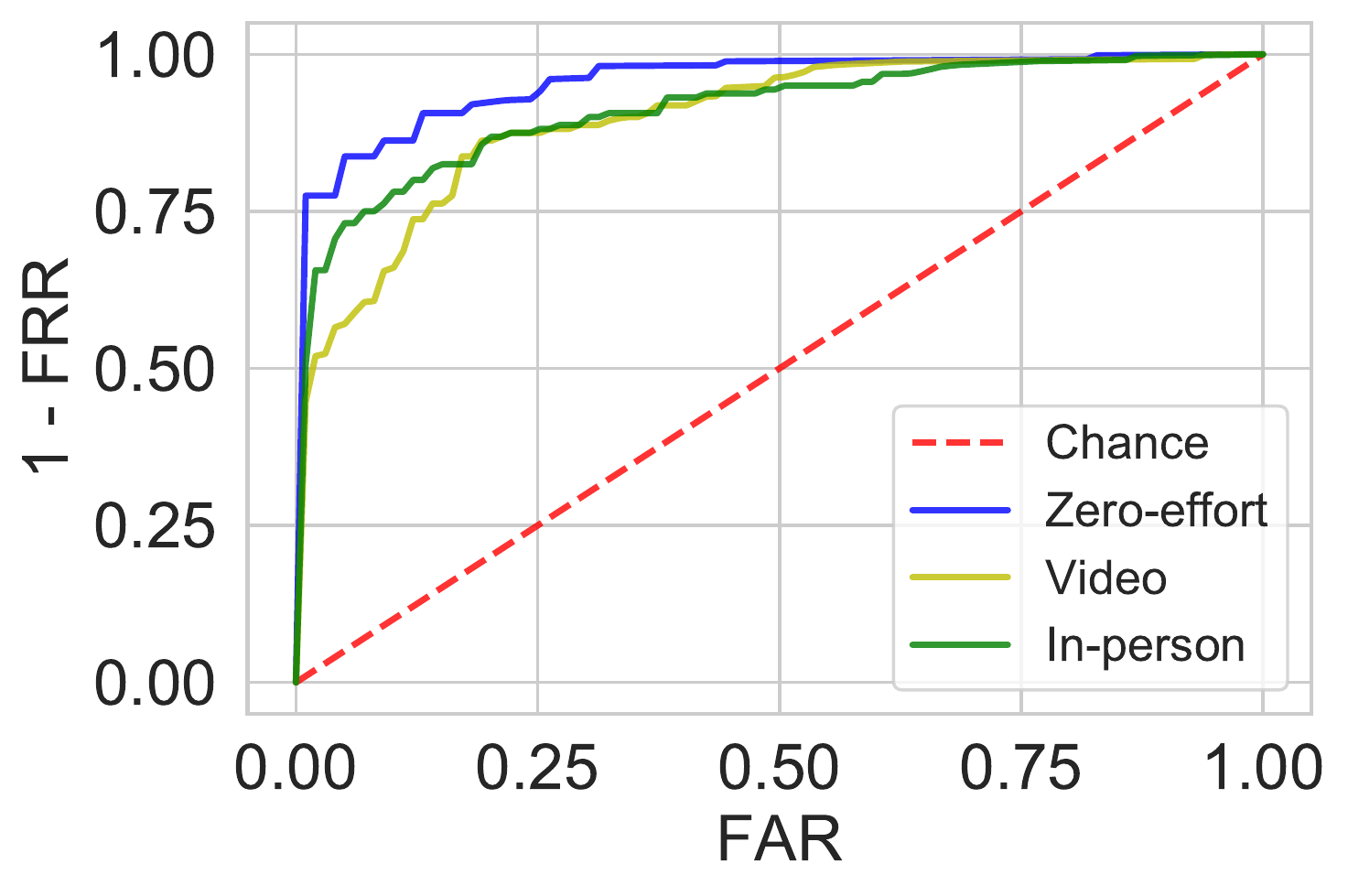}
   \label{fig:roc3}
   \caption{\onobject}
\end{subfigure}
\hfill
\begin{subfigure}[b]{0.33\textwidth}
    \includegraphics[width=\linewidth]{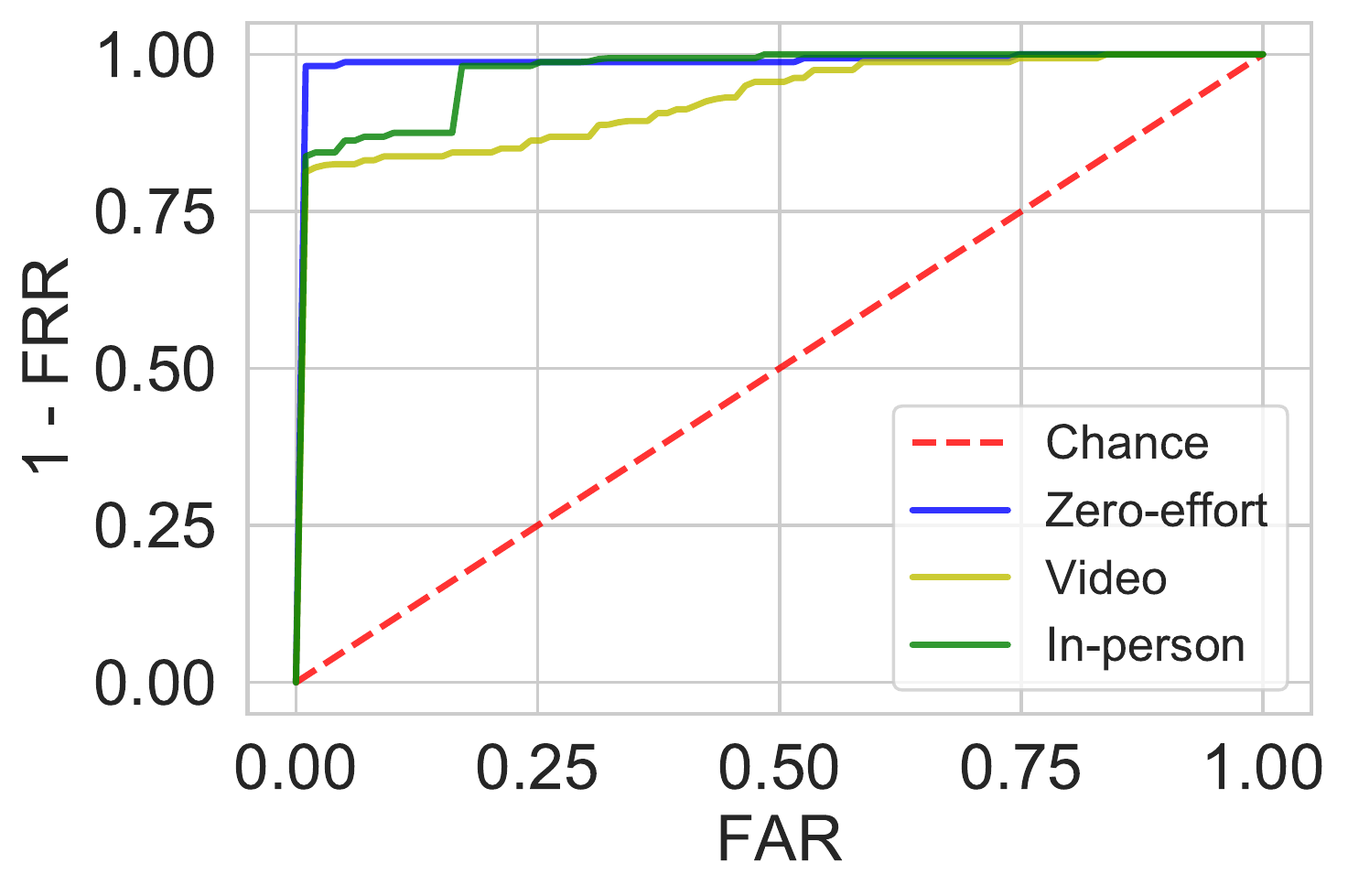}
   \label{fig:roc2}
   \caption{\offobject}
\end{subfigure}
\hfill
\begin{subfigure}[b]{0.33\textwidth}
    \includegraphics[width=\linewidth]{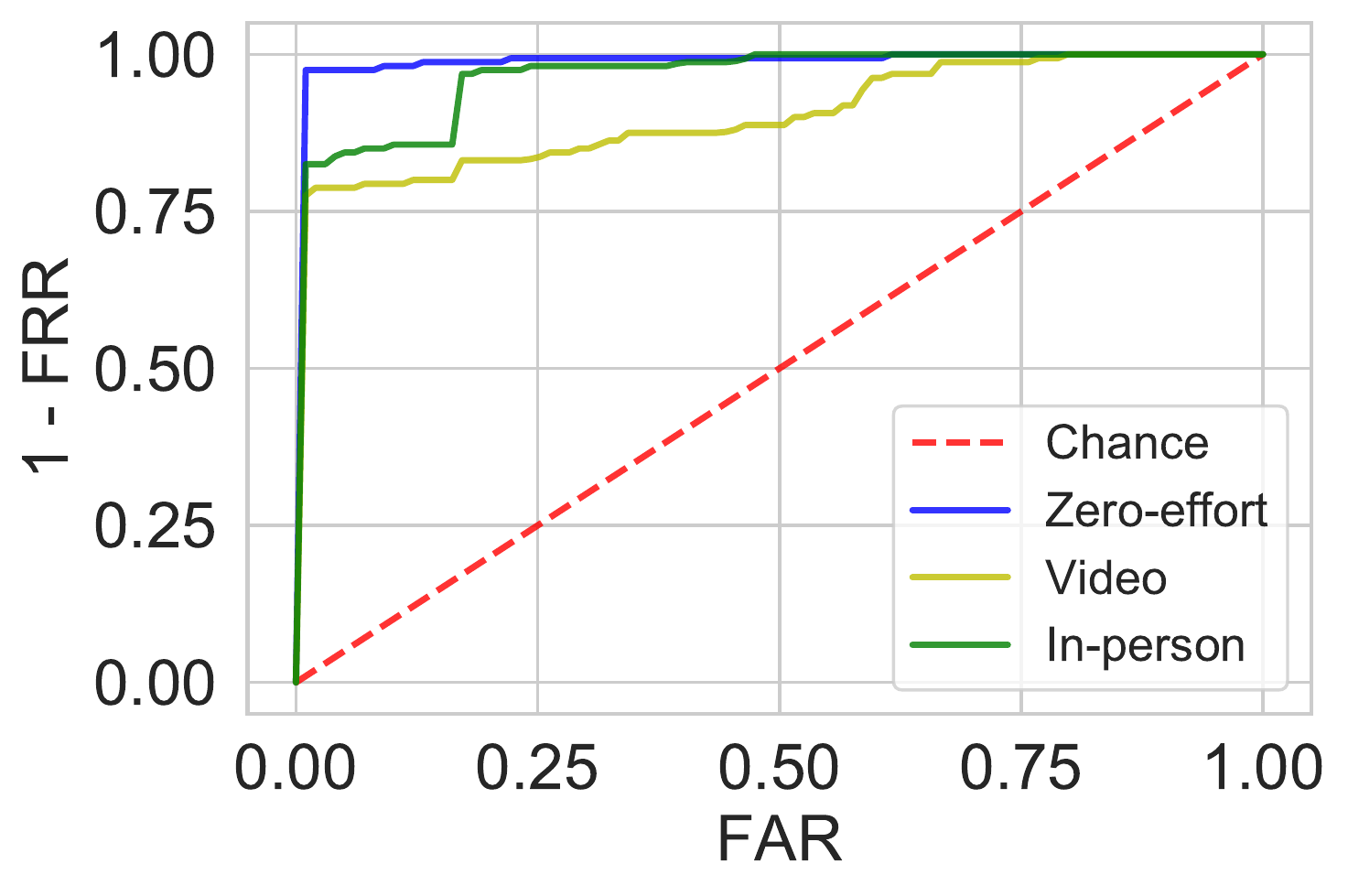}
   \label{fig:roc1}
   \caption{\combined}
\end{subfigure}

\caption{The plots above show the ROC curves for three system configurations respectively based on average FARs from single interaction types. Each curve represents a different group of attacks, i.e., zero-effort (blue), in-person (green), and video-based (yellow) attacks.}
\label{fig:roc}
\end{figure*}

First, we examine FRRs for individual smart objects that the user interacts with.
Next, we inspect the performance of ensembles of base-classifiers that are responsible for interpreting different interactions with objects.
Finally, we compare the performance of voting and stacking meta-classifiers by examining the receiver operating characteristic (ROC) curve for an ensemble of all available object interactions.

In Table~\ref{tab:frratfar}, we present FRRs at 1\% and 10\% FAR thresholds averaged across all objects for three types of attacks targeting a dedicated user.
\kk{Figure~\ref{fig:roc} shows their averaged ROC curves.}
Table~\ref{tab:frratfar_off_on} presents FRRs for individual smart objects in respect to zero-effort attacks without a dedicated victim, i.e., the results are averaged across all users being considered a victim.
\kk{For each attack, we calculate FRRs and FARs using different system configurations, including \onobject{}, \offobject{}, and \combined{}.}
For \offobject{}, only the top performing features are selected.
For each smart object, Table~\ref{tab:top_features} shows what other objects these features were extracted from.

\begin{table}[]
    \caption{For the \offobject{} configuration, we list all the co-located objects whose sensors provided the best features for individual object classifiers.}
    \label{tab:top_features}
    \resizebox{\linewidth}{!}{
\begin{tabular}{ll}
\toprule
Object           & Features from                    \\
\midrule
Right cupboard door  &  Left cupboard door \\
\addlinespace[0.5em]
\multirow{2}{*}{Wide cupboard}  & Left cupboard door, Coffee machine,\\
& Microwave door  \\
\addlinespace[0.5em]
\multirow{3}{*}{Left cupboard door}  & Right cupboard door, Wide cupboard,\\
& Oven door, Coffee machine,\\
& Pull-out drawer   \\ 
\addlinespace[0.5em]
\multirow{4}{*}{Narrow cabinet} & Right cupboard door, Wide cupboard,\\
& Left cupboard door, Oven door,\\
& Coffee machine, Pull-out drawer,\\
& Microwave door \\
\addlinespace[0.5em]
\multirow{4}{*}{Oven door} & Right cupboard door, Wide cupboard,\\
& Left cupboard door, Narrow cabinet,\\
& Coffee machine, Pull-out drawer,\\
& Microwave door \\
\addlinespace[0.5em]
\multirow{2}{*}{Coffee machine} & Right cupboard door, Wide cupboard,\\
& Pull-out drawer, Microwave door \\
\addlinespace[0.5em]
\multirow{2}{*}{Pull-out drawer} & Right cupboard door, Wide cupboard,\\
& Left cupboard door, Oven door \\
\addlinespace[0.5em]
\multirow{2}{*}{Microwave door} & Right cupboard door, Wide cupboard,\\
& Narrow cabinet, Coffee machine \\
\bottomrule
\end{tabular}
}
\end{table}

\kk{In the training phase, we only use samples collected during participants' regular interactions with the smart environment.
This is because we consider an attacker who has access to the facilities.
For example, a malicious co-worker whose typical interaction samples would be known by the system.}
A zero-effort attack, in which the attacker does not attempt to mimic the behavior of a legitimate user, is an indication of the baseline performance of the system.
Other types of attacks involve attackers who either watched the video of the victim interacting with objects or observed the victim personally.

We observe that for authentication using \offobject{} sensors, we achieve an average false reject rate of less than 3\% with a 1\% false acceptance rate for zero-effort attacks.
FRRs increase to 19\% for video-based attacks and to 15\% for in-person observation-based attacks, considering the same false acceptance rate.
This means that even when defending against strong video-based attacks, the system does not require the user to explicitly approve transactions in more than 80\% of cases, as the system can instead authenticate the user through their interactions with the smart environment.

For the FAR of 10\%, the FRR for zero-effort attacks drops to less than 1\%.
Similarly, FRRs for video-based and in-person attacks decrease to 18\% and 13\% respectively.
The \onobject{} configuration exhibits the worst performance among all of the configuration types, resulting in false reject rates of 25\% for the zero-effort attacks, 58\% and 50\% for the other types of attacks.
The \combined{} configuration guarantees better performance than \onobject{}, however, it exhibits worse performance than \offobject{} due to the inclusion of features extracted from on-device sensors.
It is noteworthy that the microwave door and the narrow cabinet classifiers perform significantly worse than others, which impacts the average scores. Since this effect is universal across users, this suggests that poorly-performing objects should be excluded by the meta-classifier.

Table~\ref{tab:frratfar_off_on} compares the performance of \onobject{} and \offobject{} configurations across all smart objects.
The narrow cabinet and the microwave door exhibit the worst FRRs in the \onobject{} configuration, resulting in false reject rates of 26\% and 30\% given a 1\% false acceptance rate for zero-effort attacks.
The FRRs drop to 0.4\% and 1\% when the model includes features extracted from co-located sensors.
Since the \offobject{} configuration exhibits the best performance, we focus on it for the remainder of this section.

\begin{figure*}[t]
\centering
\begin{subfigure}[b]{0.49\textwidth}
   \includegraphics[width=\linewidth]{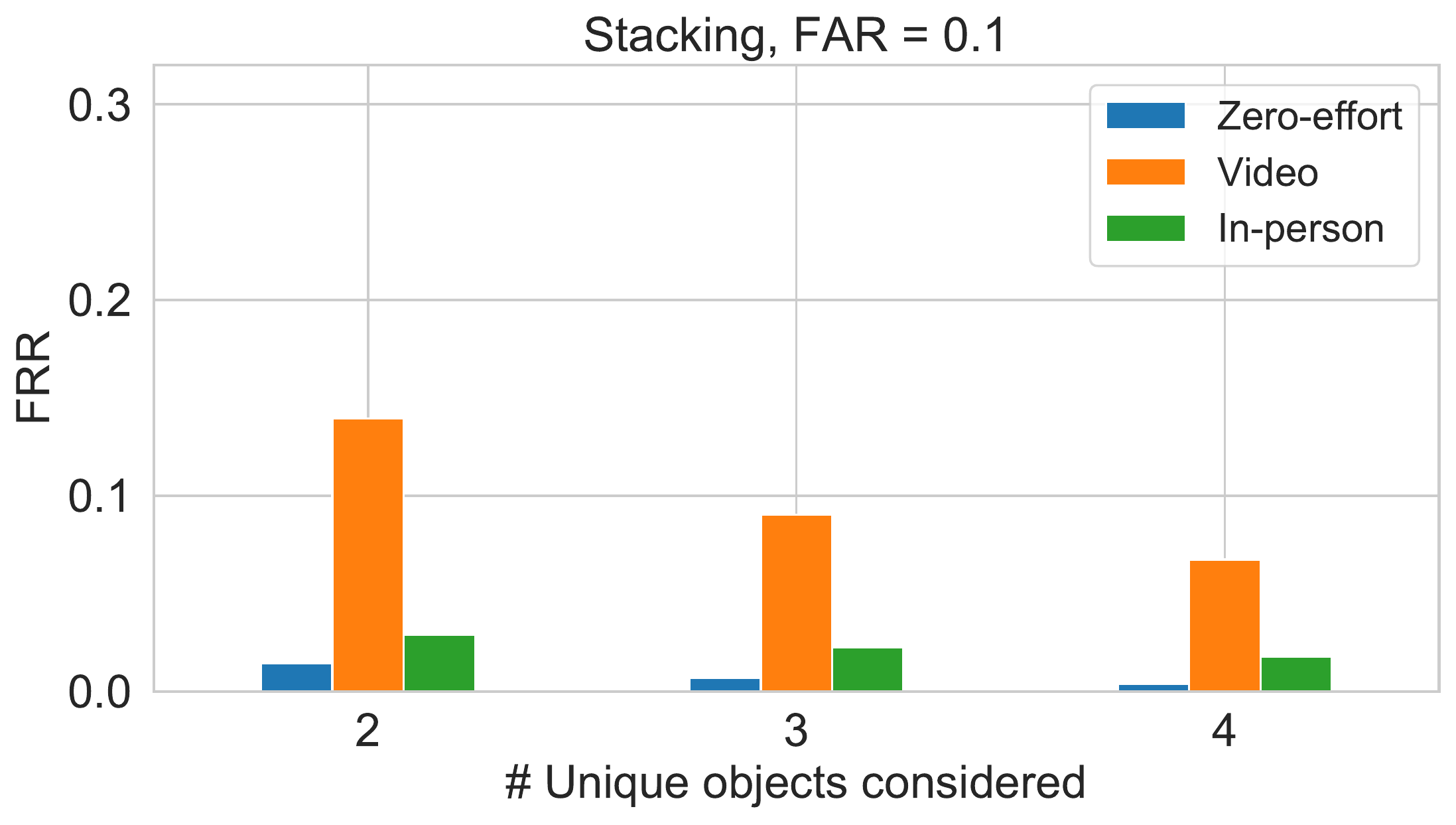}
   \label{fig:far01_stack} 
\end{subfigure}
\hfill
\begin{subfigure}[b]{0.49\textwidth}
   \includegraphics[width=\linewidth]{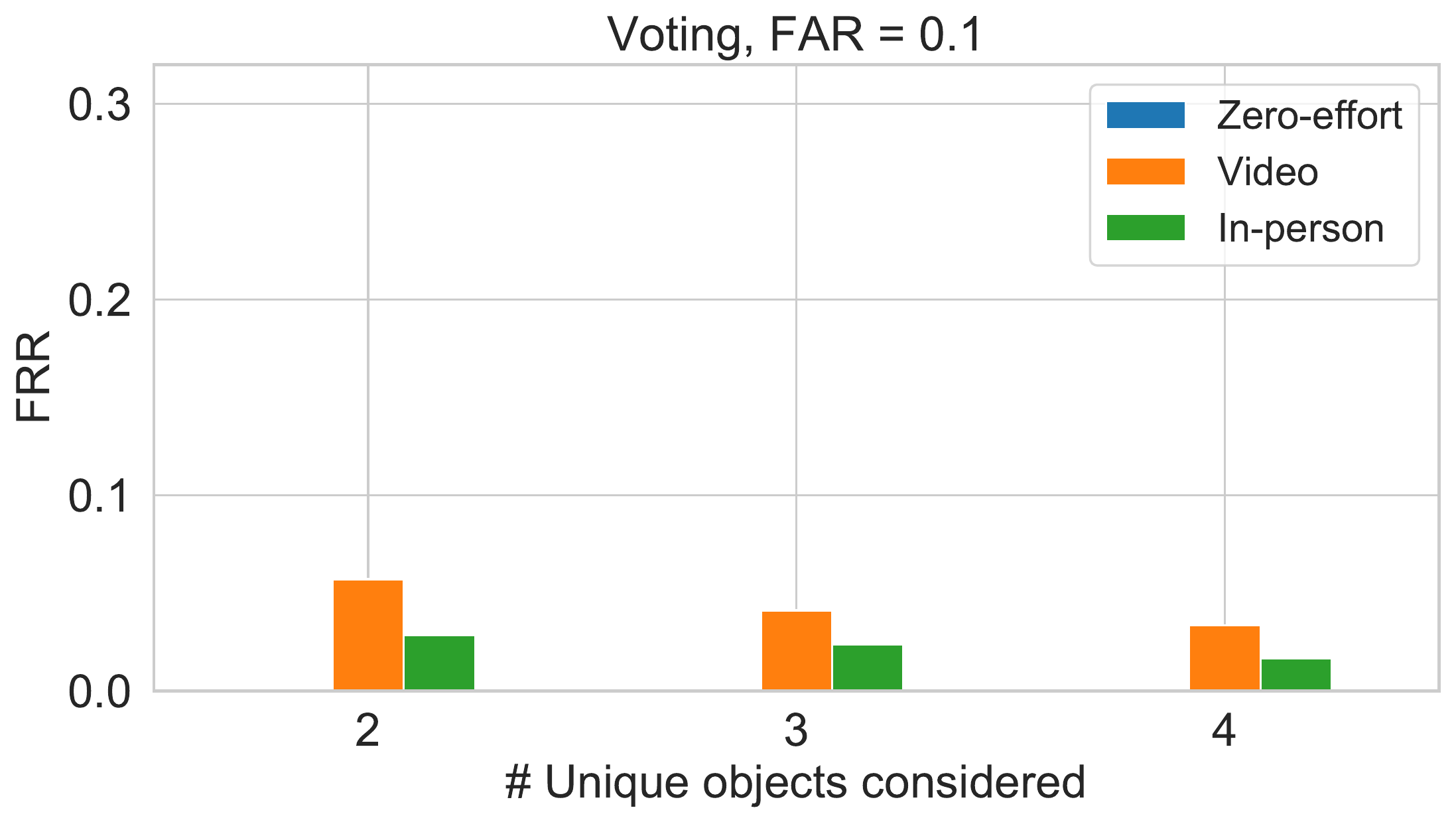}
   \label{fig:far01_vot} 
\end{subfigure}

\begin{subfigure}[b]{0.49\textwidth}
   \includegraphics[width=\linewidth]{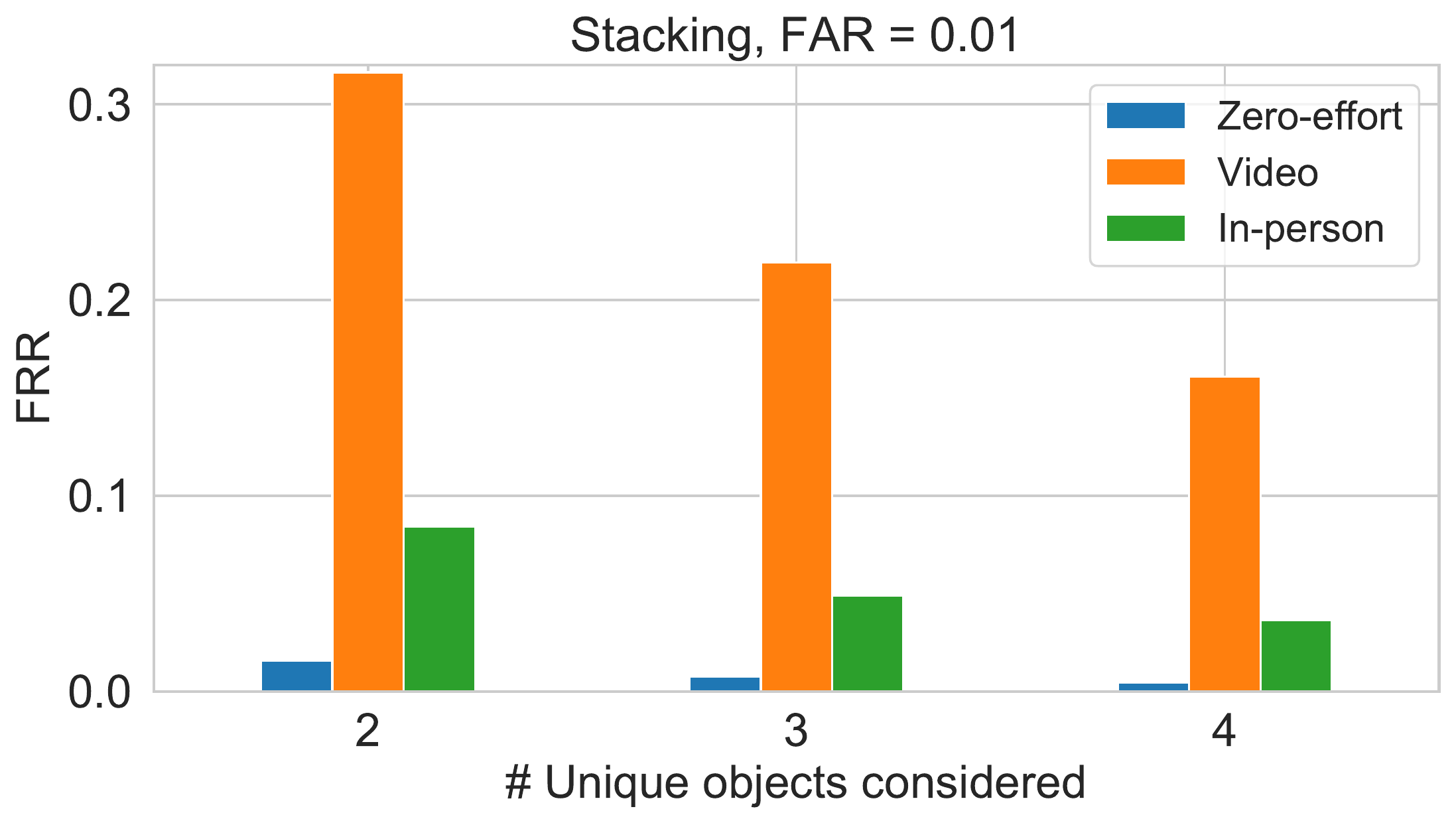}
   \label{fig:far001_stack}
\end{subfigure}
\hfill
\begin{subfigure}[b]{0.49\textwidth}
   \includegraphics[width=\linewidth]{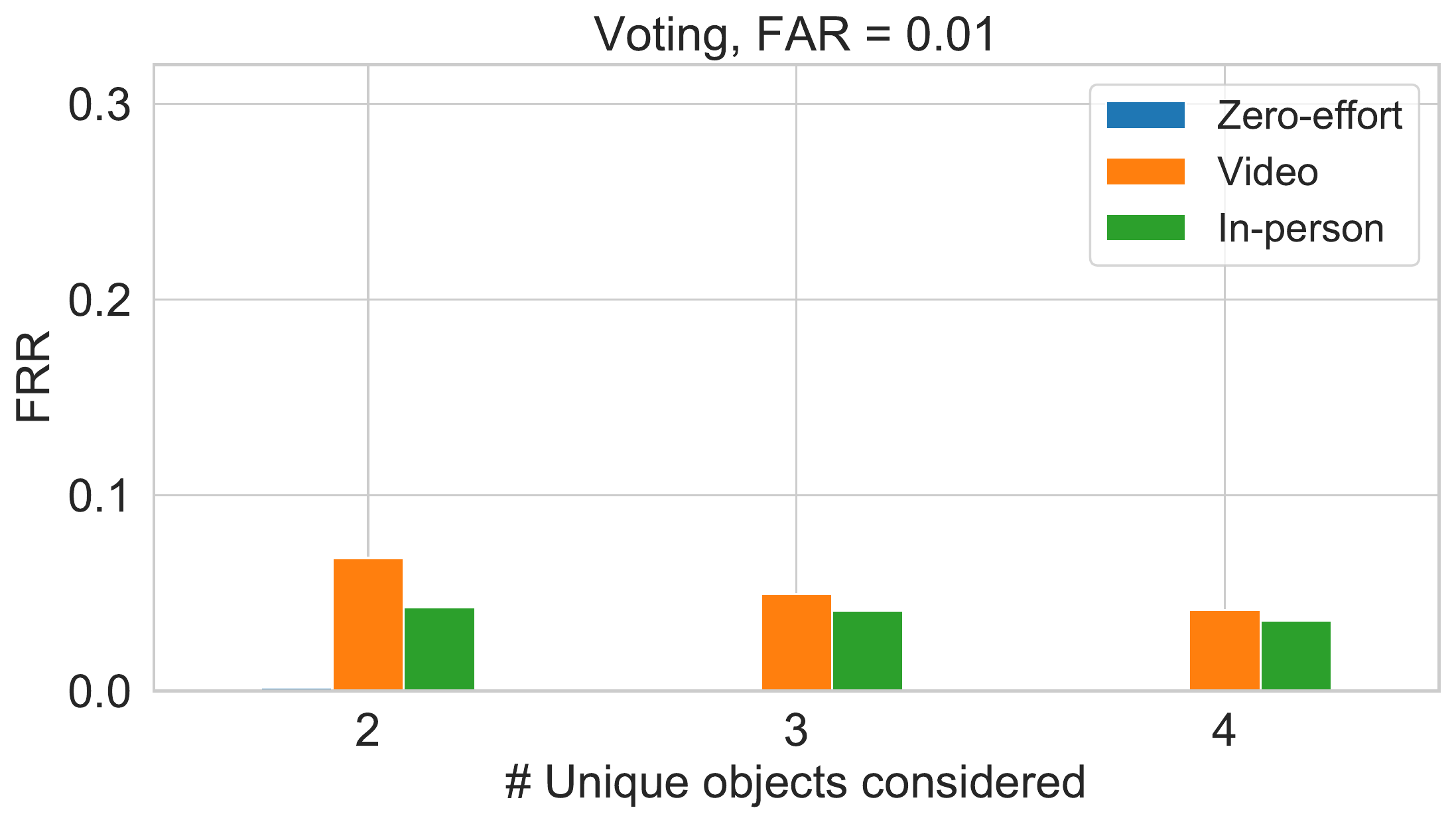}
   \label{fig:far001_vot} 
\end{subfigure}

\caption{Averaged False Reject Rates (FRRs) at different False Acceptance Rates (FARs) thresholds calculated based on the performance of different ensembles of unique objects for two meta-classifiers and the \offobject{} configuration. Each such ensemble is trained and tested separately, then the scores are averaged across the ensembles of the same type (e.g., pairs, triples of unique objects).}
\label{fig:summary_off_ens}
\end{figure*}

The attackers from the video group could watch the video of the victim performing interactions with objects as often as desired for 24 hours.
\kk{On the other hand, the attackers who observed the victim in person could follow them closely and look at the exact body and hand movements.}
To understand this phenomenon, we asked the participants to describe their strategies.
The participants from the video-based attack group watched the video three times on average before attempting to mimic the victim.
When viewing the video, participants report that they paid attention to the strength with which the victim interacted with the objects, the use of the hands (left or right), the speed of the interaction, and the body position.
The participants in the second group, on the other hand, focused mainly on the pace, strength, and rhythm of the interaction.
All attackers focused their strategy on mimicking the power and speed with which the victim interacted with objects. Additionally, most of them attempted to spend a similar amount of time per interaction as the victim did.
One of the attackers even counted the seconds spent on interactions with each object.

Considering multiple interactions with various objects can further improve the system's performance.
\kk{Figure~\ref{fig:summary_off_ens} shows averaged FRRs at two FAR thresholds of 10\% and 1\% for different ensembles of objects for the voting and stacking meta-classifiers given the \offobject{} configuration.}
We focus on the \offobject{} configuration here, as it exhibits the best performance out of the three considered configurations, and thus best demonstrates the potential performance gains that can be achieved.
This could be further improved by adjusting the weights, i.e., assigning smaller ones to interactions that exhibit worse performance.
Generally, allowing the system to consider more interactions before authenticating the user results in better performance.

Overall, the voting method outperforms the stacking meta-classifier in our scenario.
This method is also computationally less complex since it does not involve training another classifier with the predictions of the base-classifiers.
The voting meta-classifier achieves a false reject rate of less than 1\% with an FAR of 1\% whereas the stacking classifier obtains an FRR of 2\% for the zero-effort attacks.
The video-based attacks for the stacking classifier achieve an FRR of 32\% when considering the ensemble of two unique objects given an FAR of 1\%.
On the other hand, the voting classifier obtains 8\% FRR given the same FAR threshold.
This means that for the voting classifier, the system can spare the user an explicit phone-based authentication in 92\% of cases.
We included only four smart objects in this analysis but considering more unique smart objects results in further improvements of the system performance.

\section{Limitations}
\textbf{No concurrent device use. } In our experiment, we limit interactions with any device to a single user at a time. In the experiments, this was necessary to obtain accurate identity labels to establish the distinctiveness of device interactions. This limitation may lead to two potential problems in practice. If two users are interacting with different devices in the same room simultaneously or in short sequence, this may lead to decisions made using multiple device interactions to be wrong. This can be avoided by only using interactions with the target device (the device requiring authentication) to make the decision. In addition, simultaneous interactions may affect the sensor signatures and make it harder to match fingerprints for either of them. However, simultaneous interactions are easily detected and either accounted for or ignored entirely.

\textbf{Limited number of users and interactions. }Due to time considerations and the unique requirements of the ongoing Covid-19 pandemic, we could only capture device interactions in a single session. This limits our analysis for different levels of FAR and FRR, as the total number of samples and attacker/victim pairs are too low to make a statistically robust analysis of extremely low FAR levels. Given the promising results shown by our current analysis, we plan to collect an additional large-scale dataset in the future.

\textbf{Consecutive user sessions. }In our experiment, sessions for different users were conducted one after the other. In theory, it would be possible for environmental effects to be present during one user's session but not for others, thereby leading to classifiers learning these effects as a proxy for user identity. For example, a sound pressure sensor may pick up increased ambient noise during a user's session. However, the fairly strong increase in FAR caused by imitation attacks (video and in-person) suggests that the classifiers capture (somewhat imitable) true user behavior as it is unlikely users would attempt to match the original environmental conditions during their attack.

\section{Conclusion}
\label{sec:conclusions}
In this paper, we have introduced a system to authenticate users in smart environments based on naturally occurring interactions with objects around them.
Notably, our system does not require any sensors on the object itself but makes use of sensors placed arbitrarily in the room.
We have conducted an experiment in real-world settings with a total of 13 participants, which shows that using these kinds of smart object interactions for authentication is feasible.
This is a crucial finding because there is a need for stronger authorization controls in such environments, but many smart devices offer only limited interfaces to implement security features. Therefore, current systems often rely on cumbersome app-based authentication methods that require the user to always have their phone at hand.
Our system can complement such phone-based authentication methods and reduce how often a user has to explicitly approve a transaction in the smart home companion app.

We show that our system demonstrates good authentication performance against zero-effort attacks, with less than 1\% of transactions requiring external approval at an FAR of 1\% when considering a single object interaction.

When attackers attempting to imitate the victim's behavior after observing them in-person or through video footage are considered, the user has to approve more transactions explicitly to maintain a 1\% FAR.
However, the system can still authenticate more than 80\% of transactions unobtrusively when considering video-based attackers, rising to 85\% of transactions for in-person attacks.
We also show that the system's confidence in the authentication decision can be significantly improved if more than one object interaction is considered. 
\kk{Including more interactions with objects can further increase the authentication success rates to 92\% even when considering the strongest attacker.}

These promising results and the potential for easy deployment make this behavioral biometric system a good candidate to improve the security of smart environments in a seamless and unobtrusive manner.
We make our entire dataset and the code needed to reproduce our results available online to allow researchers to build on our work.

\bibliographystyle{plain}
\bibliography{main}

\end{document}